\documentclass{ws-revol}

\def\ltsim{\raise0.3ex\hbox{$<$\kern-0.75em\raise-1.1ex\hbox{$\sim$}}}
\def\gtsim{\raise0.3ex\hbox{$>$\kern-0.75em\raise-1.1ex\hbox{$\sim$}}}

\begin{document}

\chapter{Hydrodynamical description of collective flow}

\author{Pasi Huovinen}
\address{School of Physics and Astronomy, University of Minnesota,\\
         Minneapolis, MN 55455, USA}

\abstract{I review how hydrodynamical flow is related to the observed
          flow in ultrarelativistic heavy ion collisions and how
          initial conditions, equation of state and freeze-out
          temperature affect flow in hydrodynamical models.}

\tableofcontents

\newpage

\section{Introduction}
  \label{intro}

One of the goals of the experimental heavy ion program at
ultrarelativistic energies is to create a dense system of strongly
interacting particles. It is hoped that the particles formed in the
primary collisions would rescatter often enough to reach local thermal
equilibrium and behave as a particle fluid, not as a cloud of free
particles. If such a state is reached, the finally observed particles
should depict signs of collective behaviour such as flow.

Our intuitive understanding of flow, i.e.\ collective motion, is
closely tied to a classical macroscopic description of flow using the
language and tools of hydrodynamics. This means that it is often
easiest to use hydrodynamical concepts like temperature, pressure and
flow velocity to describe collective motion even if the applicability
of such concepts is far from certain. Hydrodynamical models are thus
particularly suitable to describe flow phenomena, but we have to be
careful not to confuse what is actually observed with our way
of describing observations.  An example of the limits of
hydrodynamical language is that there is no generally accepted
definition of flow in the context of heavy-ion collisions, but the
word is used in its intuitive meaning.

A rigorous definition of flow is beyond the scope of this review.
Instead I use a hydrodynamically practical definition of flow:
collective flow is correlation of position and momentum during the
dense, interacting stage of the collision regardless of the origin of
these correlations. This means that I also call flow the correlation
between the longitudinal momentum and the position of particles which
has its origin in the initial particle producing processes. In a
hydrodynamical model these correlations are manifested as an initial
non-zero longitudinal velocity field.

Unlike transverse and longitudinal flow, directed and elliptic flow do
not directly refer to collective motion but to certain emission
pattern where particle emission is not azimuthally isotropic (for
definitions see sections~\ref{v1} and~\ref{v2}). In principle elliptic
anisotropy could be entirely due to the shape of the surface of the
source and be finite even if flow the velocity is
zero~\cite{Huovinen:QM}.  Therefore to call elliptic anisotropy
elliptic flow is unfortunate but firmly established in the
literature. It has to be remembered, however, that even if elliptic
anisotropy is not necessarily a sign of collective motion, it is a
collective effect.

Hydrodynamics connects the conservation laws to the equation of state,
viscosity and heat conductivity of the fluid. Thus the properties of
matter and flow are intimately connected and we hope to learn about
the equation of state of nuclear matter by studying the flow in heavy
ion collisions. In practice, however, this is a challenging task
because of the nonlinear nature of the equations of hydrodynamics and
the many unknowns in the hydrodynamical description of heavy-ion
collision. In this review I describe briefly the basic concepts of a
hydrodynamical model and the kind of collective flow generated in
hydrodynamical simulation. My emphasis is on details which have
particular significance in the description of elliptic flow and how
initial shape of the system, equation of state and freeze-out
temperature affect elliptic flow in $Au+Au$ collisions at the
Relativistic Heavy Ion Collider (RHIC) at Brookhaven National
Laboratory (BNL) at $\sqrt{s}=130$ GeV/$A$ energy.

The use of hydrodynamics has a long tradition in heavy ion
physics. Consequently there are many previous
reviews~\cite{Stocker,Clare} and introductory
articles~\cite{Blaizot:1990,Csernai,Rischke:1998} where a reader can
find a more detailed discussion of hydrodynamical models. A
complementary view of flow and its development in hydrodynamics is
also provided in by Kolb and Heinz~\cite{Peter-Uli} in this
volume. Flow in heavy-ion collisions is discussed from the
experimental point of view in reviews by Reisdorf and
Ritter~\cite{Reisdorf:1997} and Herrmann {\em et al.}~\cite{Herrmann},
although their main emphasis is in collision energies below the
$\sqrt{s}=17.2$ GeV/$A$ collision energy of the Super Proton
Synchrotron (SPS) at CERN. These reviews also provide a discussion of
the experimental detection of flow which is beyond the scope of this
review.

\section{Hydrodynamical model}
  \label{hydro}

In a hydrodynamical description a heavy ion collision is basically
assumed to proceed as follows: In the initial collision a large amount
of the kinetic energy of the colliding nuclei is used to create a
large number of secondary particles in a small volume. These particles
will subsequently collide with each other sufficiently often to reach
a state of local thermal equilibrium. When the system has reached
local equilibrium it is characterized by the fields of temperature,
$T(x)$, chemical potentials associated with conserved charges,
$\mu_i(x)$ and flow velocity, $u^{\mu}(x)$. The evolution of these
fields is then determined by the hydrodynamical equations of motion
until the system is so dilute that the assumption of local thermal
equilibrium breaks down and the particles begin to behave as free
particles instead.

Besides the fact that we want to be able to describe the system using
a few thermodynamic variables, the hydrodynamic description has
additional advantages. Hydrodynamical models are relatively simple and
one essentially does not need any information other than the
equilibrium equation of state of nuclear matter to solve the equations
of motion. Once the equation of state and the initial state of
evolution are defined, the expansion dynamics is determined and there
is no need to know the details of the interaction on the microscopic
level. This is especially practical when one wants to study the
transition from hadronic to partonic degrees of freedom; a transition
for which the details on the microscopic level are largely
unknown. The use of familiar concepts like temperature, pressure and
flow velocity also leads to an intuitive and transparent picture of
the evolution of the collision. The price to be paid for these
advantages is a set of bold assumptions: local kinetic and chemical
equilibrium and lack of dissipation. This set of assumptions may or
may not be valid in such a small system as that formed in a heavy ion
collision.

There is no proper proof for thermalization in heavy ion
collisions. Instead, one has to revert to a heuristic comparison of
collision rates of secondary particles with the lifetime of the
collision system. At a temperature $T\gtsim 200$ MeV, for example, the
density of partons is $n\gtsim 4$~fm$^{-3}$ in a two flavor
plasma. When the cross section is approximated by the perturbative QCD
(pQCD) gluon-qluon scattering cross section of $\sigma_{gg\rightarrow
gg} \approx 3$ mb, the average mean free path, $\lambda = 1/\sigma n
\approx 0.8$ fm, and the time between two collisions is an order of
magnitude smaller than hydrodynamically estimated lifetime of the
system, $\tau = 10$--20 fm. Thus the partons should scatter several
times and the system has a chance to thermalize. Another way to argue
for thermalization in heavy ion collisions is simply to refer to the
elliptic flow data which can be reproduced by a hydrodynamical model,
but not as well by transport models~\cite{Heinz:2001}.

  \subsection{Basics}

The equations of motion of relativistic fluid dynamics are the expressions
for local conservation of energy and momentum and any conserved charge:
\begin{equation}
   \partial_{\mu}T^{\mu\nu} = 0 {\hspace{1cm}} {\rm and} {\hspace{1cm}}
   \partial_{\mu}j^{\mu}_i = 0,
\end{equation}
where $T^{\mu\nu}$ is energy momentum tensor and $j_i^\mu$,
$i=1,\ldots,n$ are the four-currents of conserved charges. Without any
additional constraints these $4+n$ ($n$ is the number of conserved
currents) equations contain $10+4n$ unknown variables. The simplest
and most commonly used approach to close this system of equations is
the ideal fluid approximation which reduces the number of unknown
variables to $5+n$.

In the ideal fluid approximation the energy momentum tensor
\begin{equation}
    T^{\mu\nu} = \int \frac{\mathrm{d}^3{\bf p}}{(2\pi)^3 E}\,
                                  p^\mu p^\nu f(x,{\bf p}),
\end{equation}
and currents $j_i$ are supposed to have forms
\begin{equation}
    T^{\mu\nu} = (\epsilon + p) u^\mu u^\nu - pg^{\mu\nu}
    {\hspace{1cm}} {\rm and} {\hspace{1cm}}
    j^{\mu}_i = n_i u^\mu,
\end{equation}
where $\epsilon$, $p$ and $n_i$ are energy density, pressure and
number density of charge $i$ in the local rest frame of the fluid and
$u^\mu$ is the flow four-velocity of the fluid. In other words all
dissipative effects such as viscosity and heat conductivity are assumed
to be zero and the fluid is always in perfect
local kinetic equilibrium. The additional equation needed to close the
system of equations is provided by the equilibrium equation of
state (EoS) of the matter, which connects the pressure to the
densities: $P=P(\epsilon,n_1,\ldots,n_n)$.

In principle it is possible to include small deviations from local
thermal equilibrium by including dissipative effects, but in practice
relativistic viscous hydrodynamics is very difficult to implement and
has not yet been done~\cite{Rischke:1998}. For estimates of the effects
of viscosity, see refs.~\cite{Muronga:2001,Teaney:2003}.

The numerical solution of the hydrodynamical equations of motion in
all three spatial dimensions is a tedious problem. In most approaches
some approximate symmetry is applied to effectively reduce the number
of spatial dimensions to two or one. A trivial simplification is to
assume cylindrical symmetry in the description of central collisions
of spherical nuclei.

Another popular approximation is the Bjorken model~\cite{Bjorken}
where the longitudinal flow is assumed to be of the scaling form
$v_z = z/t$ at all times. This requirement leads to boost invariance
of the system: its pressure and energy density do not depend on the
longitudinal coordinate $z$, if compared at the same proper time
$\tau =\sqrt{t^2-z^2}$. The solution of the equations of motion becomes
independent of boosts along the beam axis and it is sufficient to
solve the equations of motion in the transverse plane at $z=0$. The
obvious drawback in this approximation is that the results are
independent of rapidity and one is limited to discuss only transverse
behaviour. With the exception of results by Hirano {\em et
al.}~\cite{Hirano:2001,Hirano:2002}, the results discussed in this
review are obtained in boost invariant calculations. At RHIC energies
the transverse flow results at mid rapidity are very similar in both
boost invariant and non-boost invariant calculations.

  \subsection{Initialization}
    \label{init}

Local thermal equilibrium is one of the assumptions of a
hydrodynamical model, but the model itself does not specify the
mechanism that leads to an equilibrated state. Since at RHIC energies
the initial particle production is definitely not an adiabatic
process, hydrodynamics cannot be used to describe the initial
collision, but the hydrodynamical evolution must begin at a sufficient
time after the initial collision when the system has had time to reach
thermal equilibrium. The initial state of the system, i.e. the density
distributions and flow velocities at the beginning of the hydrodynamic
evolution, are not given by the model either but must be given as
external input.

When analysing flow it is not enough to know what the maximum density
or temperature reached in collision was. The maximum flow velocity
before freeze-out depends also on pressure gradient close to the edge
of the system. This makes particle distributions at high $p_T$
sensitive, not only to the maximum initial pressure, but also to the
initial density profile~\cite{Dumitru:1998}. The elliptic flow, on the
other hand, is closely related to the initial asymmetry of the
system~\cite{Ollitrault:1992,Kolb:2001b}. This makes choosing the
initial distributions an essential part of modelling flow.

The simplest method to determine the initial state is the one proposed
by Hwa and Kajantie~\cite{Hwa:1985}: Since ideal fluid expansion is
isentropic and entropy is related one-to-one to the measured
multiplicity at a fixed freeze-out temperature and chemical potential,
the final multiplicity gives also the initial entropy of the
system. Only a choice of the initial size of the system is needed to fix
the (average) initial entropy density and if the equation of state is
known, all other thermodynamical properties follow. This approach does
not tell anything about the initial density distributions and more
constraints are needed to study flow.

In boost-invariant expansion it is sufficient to specify the density
profiles in the transverse plane. A plausible approach is to localize
the wounded nucleon model~\cite{WNM} and assume that the density in
the transverse plane is proportional to the number of participants per
unit area in the transverse plane. For two nuclei colliding with
impact parameter $\mathbf{b}$, the density of participants can be
calculated from a geometric formula (ref.~\cite{Blaizot:1990} and
references therein):
\begin{eqnarray}
 n_{\mathrm{WN}}(\mathbf{s};\mathbf{b})
 & = & T_A(\mathbf{s}+{\textstyle \frac{1}{2}}\mathbf{b})
  \left[1-\left(1-\frac{\sigma_{pp} T_B(\mathbf{s}-\frac{1}{2}\mathbf{b})}{B}\right)^{\!B}\right]
 \nonumber \\
 & + & T_B(\mathbf{s}-{\textstyle \frac{1}{2}}\mathbf{b})
 \left[1-\left(1-\frac{\sigma_{pp} T_A(\mathbf{s}+\frac{1}{2}\mathbf{b})}{A}\right)^{\!A}\right],
  \label{wounded}
\end{eqnarray}
where $\sigma_{pp}$ is the inelastic proton-proton cross section at the
collision energy, $T_A$ is the nuclear thickness function of nucleus
$A$;
\begin{equation}
  T_A(\mathbf{s}) = \int_{-\infty}^{\infty} \mathrm{d}z\,\rho_A(\mathbf{s},z),
  \label{TA}
\end{equation}
and $\rho_A(\mathbf{r})$ is nuclear density given by a Woods-Saxon
distribution.

At SPS, the final particle multiplicity scales with the number of
participants~\cite{WA98}. Thus it is natural to assume that the
initial entropy density scales with the number of 
participants~\cite{Ollitrault:1992,Teaney:2001,Teaney:2002} and the
initial entropy density distribution is given by
\begin{equation}
  s(\mathbf{s};\tau_0;\mathbf{b})
   = K_s(\tau_0)n_{\mathrm{WN}}(\mathbf{s};\mathbf{b}),
\end{equation}
where $K_s(\tau_0)$ is a proportionality constant chosen to reproduce
the observed final particle multiplicity in central collisions. This
constant depends on the initial time of hydrodynamical evolution,
$\tau_0$, which must be chosen separately. In the following this
parametrization is called sWN, as in ref.~\cite{Kolb:2001b}.

Strictly speaking the hydrodynamical approach does not allow both the
transverse energy $\mathrm{d}E_T/\mathrm{d}y$ and the multiplicity
$\mathrm{d}N/\mathrm{d}y$ to scale with the number of participants. The
data, however, does not rule out the possibility that transverse
energy is proportional to the number of participants~\cite{WA98}. Thus
one can make an assumption~\cite{Kolb:2000a,Kolb:2000b,Huovinen:2001}
that the initial energy density, not the initial entropy density, scales
with the number of participants as given in Eq.~(\ref{wounded}):
\begin{equation}
  \epsilon(\mathbf{s};\tau_0;\mathbf{b}) 
    = K_{\epsilon}(\tau_0)n_{\mathrm{WN}}(\mathbf{s};\mathbf{b}),
\end{equation}
with a proportionality constant $K_{\epsilon} \neq K_{s}$.
By analogy to the label sWN, this initialization is labeled as eWN.

With increasing collision energy one expects that the hard collisions
between incoming partons become more and more important and finally
dominate particle production~\cite{KLL87}. In that limit each
nucleon-nucleon collision contributes equally to the particle and
energy production and the number of produced particles scales like the
number of binary nucleon-nucleon collisions in the transverse
plane. It is given in terms of nuclear thickness functions (Eq.~(\ref{TA}))
as
\begin{equation}
 n_{\mathrm{BC}}(\mathbf{s};\mathbf{b})
  = \sigma_{pp}\,T_A(\mathbf{s}+{\textstyle \frac{1}{2}}\mathbf{b})\,
            T_B(\mathbf{s}-{\textstyle \frac{1}{2}}\mathbf{b}).
\end{equation}
If the system thermalizes quickly via inelastic collisions the density
of produced particles defines the initial entropy density at the
beginning of the hydrodynamical expansion. Thus the initial entropy
density should be proportional to the number of binary collisions:
\begin{equation}
  s(\mathbf{s};\tau_0;\mathbf{b})
    = \kappa_s(\tau_0)n_{\mathrm{BC}}(\mathbf{s};\mathbf{b}),
\end{equation}
which defines parametrization sBC for initialization. The
proportionality constant is now labeled $\kappa$ to emphasize that its
value is different from the values of $K_s$ and $K_{\epsilon}$ in the
sWN and eWN parametrizations.

\begin{figure}
  \begin{center}
    \psfig{file=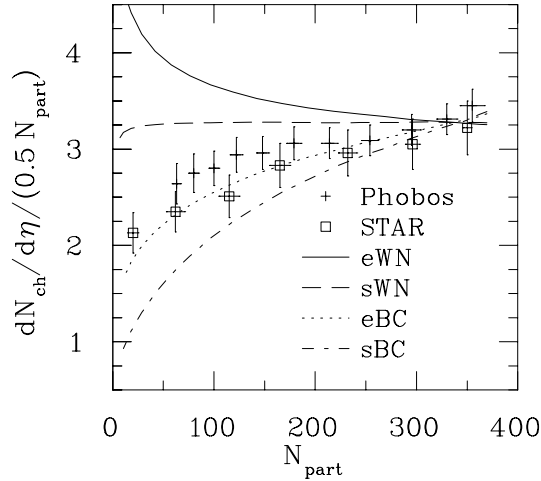,width=7cm}
  \end{center}
  \caption{Charged particle yield per participating nucleon pair at
           midrapidity as a function of the number of participants
           for different initialization models discussed in the
           text\protect\cite{Kolb:2001b}. All curves were normalized to
           $dN_{ch}/d\eta = 550$ for 5\% of the most central collisions
           ($b=2.3$ fm.) The data are from
           refs.\protect\cite{Phobos-mult,Star-mult}.}
  \label{dN-Npart}
\end{figure}

One can also argue that each binary collision contributes equally not
only to particle production but also to the energy carried by the
produced particles. In that case the initial energy density should be
proportional to the number of binary
collisions~\cite{Hirano:2001,Hirano:2002}:
\begin{equation}
  \epsilon(\mathbf{s};\tau_0;\mathbf{b})
    = \kappa_\epsilon(\tau_0)n_{\mathrm{BC}}(\mathbf{s};\mathbf{b}).
\end{equation}
This parametrization is called eBC.

Each of these parametrizations leads to a different centrality
dependence of the multiplicity. Thus one can use data to differentiate
between them without any reference to flow. As shown in
fig.~\ref{dN-Npart} the multiplicity data shows slightly stronger than
linear dependence on the number of
participants~\cite{Phobos-mult,Star-mult,Phenix}. Parametrization eBC
is closest to the data, but the linear behaviour of sWN is not far
from the data either. Even if the multiplicity data constrains the
initial parametrization somewhat, there is still freedom in choosing a
combination of these parametrizations. One should also remember that
these parametrizations are not the only possibilities.

Besides the initial distributions one has to choose the initial time
of hydrodynamic evolution, $\tau_0$. As there is no method to
calculate whether the system thermalizes, there is no method to
calculate when the system is sufficiently thermalized for the
hydrodynamic evolution to begin. Thus the initial time is another free
parameter to be chosen to fit the data or by other arguments like
saturation scale in pQCD calculations. In simulations of $Au+Au$
collisions at RHIC ($\sqrt{s}=130$ GeV), the initial time has varied from
$\tau_0=0.2$ to 1 fm~\cite{nantes}.

If the assumption of boost invariance is relaxed, the choice of
initial state becomes considerably more complicated. There are few
constraints for the longitudinal flow velocity profile or the
longitudinal density distributions. Thus the choice of a particular
parametrization and the values of the parameters is largely based on
trial and error -- tuning the model until a reasonable fit to
experimental rapidity distributions is achieved. For a sample of
initial profiles used successfully to describe longitudinal expansion
at the SPS see refs.~\cite{Ornik:1996,Sollfrank:1998,Morita} and at
RHIC refs.~\cite{Hirano:2001,Morita}. It is also instructive to
remember that even for the same EoS there are several possible initial
states which lead to an acceptable reproduction of the
data~\cite{Huovinen:1998}.

An alternative approach to determine the initial state is to use some
other model to calculate it. For example event generator~\cite{Schlei:1998}
or perturbative QCD (pQCD) calculations~\cite{Eskola:2001,JKL-ryhma}
have been used for this purpose. Even if these approaches increase the
predictive power of hydrodynamics, thermalization is still an
additional assumption.

  \subsection{Equation of State}

With the notable exception of ref.~\cite{Zschiesche} where the
equation of state (EoS) is based on chiral SU(3) $\sigma -\omega$ model
\footnote{In this paper only HBT radii were discussed and it is thus
beyond the scope of this review.} all the hydrodynamical calculations
at RHIC energies have used equations of state based on similar
structure: a hadronic phase which is constructed as a gas of free
hadrons and resonances, a plasma phase of ideal, massless partons with
a bag constant and a first order phase transition between these two
phases.

Different choices of the number of resonances included in the
ha\-dron\-ic phase, of the latent heat and of the phase transition
temperature cause minor differences in the equations of state of
different practitioners, but the major difference is whether the
hadron phase is assumed to be in chemical equilibrium or not. Thermal
models used to fit final state particle abundancies give larger
freeze-out temperatures $T_{ch}\sim 160$ MeV~\cite{Cleymans} than
kinetic freeze-out temperatures $T_f\sim 120$ MeV obtained from fits
to particle $p_T$ spectra. This discrepancy can be explained by
different cross sections for elastic and inelastic collisions. Thermal
models assume chemical equilibrium which requires frequent inelastic
collisions which change particle number. On the other hand, frequent
inelastic collisions are sufficient to maintain kinetic
equilibrium. Since the cross sections for particle number changing
collisions are much smaller than for collisions where particle number
does not change (elastic and quasi-elastic collisions), it is natural
to assume that inelastic collisions cease first and chemical
freeze-out occurs at a higher temperature than kinetic freeze-out. Thus
the system may be in local kinetic, but not chemical, equilibrium at
the later stages of its evolution.

Chemical non-equilibrium can be incorporated in the hydrodynamical
description by treating lowest lying hadron states as stable
particles~\cite{Bebie,Teaney:2002}. The particle number of each of
these hadrons forms a conserved current and a finite chemical
potential for each hadron is built up. Baryon and antibaryon chemical
potentials are also independent in this approach. Chemical
non-equilibrium changes the space-time evolution of the system only
slightly because the relation between pressure and energy density is
very similar both in chemical equilibrium and
non-equilibrium~\cite{Teaney:2002}.  The main difference is that in
chemical non-equilibrium the temperature decreases faster as energy
density decreases and thus the system reaches its freeze-out
temperature faster. How this changes the observed anisotropy will be
discussed in section~\ref{data}.

  \subsection{Freeze-out}
     \label{decouple}

At some point in the evolution the particles will begin to behave as
free particles instead of a fluid and the hydrodynamical description
must break down. When and where that happens is not given by
hydrodynamics but must be included as external input. The
conventional approach is to assume this to take place as a sudden
transition from local thermal equilibrium to free streaming particles
when the expansion rate of the system is larger than the collision
rate between particles or the mean free path of the particles becomes
larger than the system size. Finding where these conditions are
fulfilled is a nontrivial problem. Since the scattering rate is
strongly dependent on temperature the usual approximation assumes that
the freeze-out takes place on a hypersurface where temperature (or
energy density) has a chosen freeze-out value. This temperature is of
the order of the pion mass, but its exact value is largely a free
parameter which can be chosen to fit the data. In $Pb+Pb$ collisions at
the SPS ($\sqrt{s} = 17$ GeV/$A$) the calculated values of freeze-out
temperatures vary between 100 and 140 MeV~\cite{Rischke:2001}. In
$Au+Au$ collisions at RHIC ($\sqrt{s}=130$ or 200 GeV/$A$) the span of
suggested freeze-out temperatures is even wider, from
100~\cite{Adler:2001,Kolb:2002} to 160--165 MeV~\cite{JKL-ryhma,W&W}.

After choosing the surface where the freeze-out takes place, the
thermodynamic variables characterizing the state of the fluid
must be converted to spectra of observable particles. A
practical way of doing this is the Cooper-Frye
algorithm~\cite{Cooper-Frye} where the
invariant momentum distribution of a hadron $h$ is given by
\begin{equation}
  E \frac{dN}{d^{3}p} 
    =  \frac{g_h}{(2 \pi)^{3}} \int_{\sigma_{f}} 
          \frac{1}{exp[(p_{\mu}u^{\mu}-\mu)/T] \pm 1} \,
                           p^{\mu} d\sigma_{\mu},
	\label{CF}
\end{equation}
where the temperature $T(x)$, chemical potential $\mu(x)$ and flow
velocity $u^{\mu}(x)$ are the corresponding values on the decoupling
surface $\sigma_{f}$. Besides its relative simplicity, this approach
has the advantage that if the same equation of state is used on both
sides of decoupling surface, both energy and momentum are conserved.

However, the Cooper-Frye formula has a conceptual problem. For those
areas where the freeze-out surface is spacelike, the product $p^{\mu}
d\sigma_{\mu}$ may be either positive or negative, depending on the
value and direction of $p^{\mu}$. In other words, the number of
particles freezing out on some parts of the freeze-out surface may be
negative. These negative contributions are small (a few per
cent~\cite{Teaney:2001}) and are usually ignored. More refined procedures
without negative contributions have been suggested~\cite{Bugaev} but
their implementation is complicated. So far there has been no
full-fledged calculation using these procedures.

Another way to refine the hydrodynamical freeze-out procedure is to
circumvent the entire problem and switch from a hydrodynamical to a
microscopic transport model description well within the region where
hydrodynamics is supposed to be applicable~\cite{Teaney:2001,Bass}.
Besides giving a better description of freeze-out, such models include
the separate chemical and kinetic freeze-outs. The main drawback of
such models is -- besides the increased complexity -- that the correct
region where the switch from the hydrodynamical to the transport
description should take place is as uncertain as the kinetic
freeze-out surface in ordinary hydrodynamical calculation. The
educated guess employed in both refs.~\cite{Teaney:2001,Bass} is that
the switch happens immediately after hadronization.

\section{Transverse flow and its anisotropies}
\label{flow}

In hydrodynamics, flow is generated by pressure gradients. In the initial
state of a relativistic heavy-ion collision, there is only one type
of collective motion, the coherent longitudinal motion of the two
approaching nuclei. At center-of-mass energies above about 5\,GeV/$A$
nucleon pair, this initial motion can no longer be completely stopped, 
even for large nuclei ($A\geq 200$) undergoing fully central collisions,    
and a certain fraction of the collective motion in the final state is 
simply a remnant of the initial beam motion. To separate it from 
hydrodynamically generated longitudinal flow is notoriously difficult,
and therefore most of the attention focuses on {\em transverse} 
collective flow in the directions perpendicular to the beam which was
entirely absent before the collision and thus can be clearly associated
with collective behaviour generated during the collision.

One distinguishes between transverse flow in general and its
anisot\-ro\-pies. In some works in literature, the azimuthally
averaged transverse flow is called radial flow (e.g.\ ref.~\cite{Peter-Uli}).
However, in studies of heavy ion collisions at energies around 1 GeV/$A$,
radial flow is understood to mean three dimensional spherically
symmetric flow~\cite{Reisdorf:1997,Herrmann}. Therefore I prefer to
call collective motion in the transverse plane transverse flow without any
reference to its azimuthal structure. Transverse flow will be
discussed in Section~\ref{pt}.

In central ($b=0$) collisions between spherical nuclei the transverse
pressure gradients are independent of the azimuthal angle and
transverse flow field is azimuthally isotropic. On the other hand in
non-central ($b\ne0$) collisions between spherical nuclei, or in
central collisions between deformed nuclei (e.g.\ U+U), the nuclear
overlap region is initially spatially deformed in the plane transverse
to the beam, resulting in azimuthal anisotropies of the pressure
gradients and of the final collective flow pattern generated by them.
Two specific forms of anisotropy of the transverse flow field are the
``directed'' and the ``elliptic'' ``flows'' discussed in
Sections~\ref{v1} and~\ref{v2} below. They are characterized and
measured by the first two higher order expansion coefficients in an
azimuthal Fourier expansion of the final momentum spectra. These
coefficients vary for different hadron species, but the hydrodynamic
model relates their magnitudes in specific ways which depend on the
initial and freeze-out conditions and the equation of state of
the expanding matter.

\subsection{Transverse flow}
\label{pt}

In one-fluid hydrodynamics, the concept of transverse flow is
particularly simple. The pressure gradient between the dense center of
the system and the ambient vacuum causes the system to expand and
transverse velocity $v_r$ of the fluid is built up. However, if one
defines the system size by the location of the freeze-out surface, the
transverse flow velocity is {\em not} the expansion velocity of the
system. Since the system dilutes, freeze-out surface may move inwards
even if the particle fluid flows rapidly outwards. In such a case the
whole concept of expansion velocity is ambiguous.

The equation of state (EoS) is closely related to the buildup of flow:
the stiffer the EoS, the larger the flow. Unfortunately many other
factors affect the flow as well and therefore $p_T$ spectra of
particles constrain the EoS only weakly. Extreme cases like the ideal
pion gas EoS can be excluded as too
stiff~\cite{Schnedermann:1994,Sollfrank:1996}, but even the effect of
a phase transition can be compensated by changes in the initial
density and freeze-out temperature~\cite{Huovinen:1998}. At large
values of the transverse momentum, $p_T > 2$--3 GeV/$c$, the particle
distributions are also increasingly sensitive to the details of the
velocity profile and thus to the initial pressure profile. Similar
calculations may reproduce the data or not depending on the choice of
the initial profile~\cite{Dumitru:1998}.

\begin{figure}
  \begin{center}
     \psfig{file=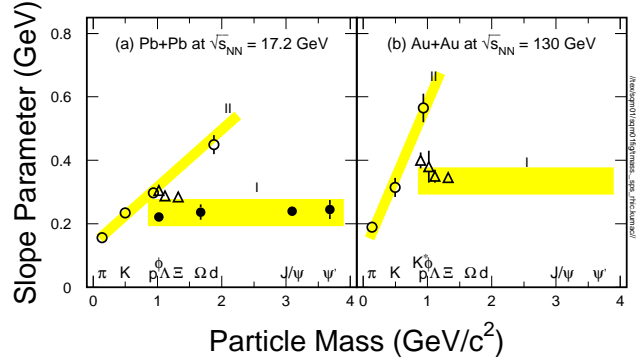,width=10cm}
  \end{center}
 \caption{Slope parameters as a function of particle mass for (a) Pb+Pb
          central collisions at the SPS ($\sqrt{s}=17.2$ GeV/$A$)
          and (b) Au+Au central collisions at RHIC
          ($\sqrt{s}=130$ GeV/$A$). From
           ref.\protect\cite{NuXu}.}
 \label{Tslope}
\end{figure}

The experimental detection of transverse flow is much more difficult
than its hydrodynamical description. Since it is not possible to
detect where each particle was emitted, it is not possible to
reconstruct the flow (or its absence) either. Instead one has to
deduce the presence of flow indirectly by comparing $p_T$
distributions of various particle species. In the experimental
literature the following procedure is often used to argue for
collective flow: To characterize the slope of the $p_T$ distribution
the transverse mass spectra at midrapidity can be fitted to a simple
Boltzmann distribution
\begin{equation}
	\frac{\mathrm{d}N}{\mathrm{d}y\,m_T\,\mathrm{d}m_T}
	\propto \exp\left( -\frac{m_T}{T_\mathrm{slope}}\right).
	\label{fit_slope}
\end{equation}
Here $T_\mathrm{slope}$ is the inverse slope parameter, often
interpreted as the apparent temperature of the source. In $p+p$
collisions at 450 GeV/c~\cite{Alper} the spectra of different
particles (e.g. $\pi$, K, p) have a characteristic inverse slope
parameter of about 140 MeV. On the other hand, as shown in
Fig.~\ref{Tslope}, in $Pb+Pb$ collisions at the SPS~\cite{NA44} and in
$Au+Au$ collisions at RHIC~\cite{NuXu} the slope parameter of $\pi$,
$K$ and $p$ increases with particle mass and collision energy. The
linear increase in particle mass has been interpreted as a sign that
$T_\mathrm{slope}$ consists of two components: (a) thermal part,
$T_\mathrm{therm}$, associated with random motion, and (b) a part
resembling collective motion with average transverse flow velocity
$\langle v_r\rangle$. Both contributions can be added and give rise to
the slope parameter~\cite{NA44}
\begin{equation}
	T_\mathrm{slope} = T_\mathrm{therm} + m\langle v_r\rangle^2.
  \label{Tsimple}
\end{equation}

However, one must warn against a too simpleminded use of this procedure
and interpretation of Eq.~(\ref{Tsimple}). First of all the
experimental distributions as well as the distributions given by
Eqs.~(\ref{CF},\ref{blastwave}) are invariant distributions whereas the
thermal distribution used in the fits of Eq.~(\ref{fit_slope}) is
not. The $m_T$ factor required to make the distribution invariant and
the integrations over the source leading to Bessel functions instead
of exponentials cause the value of the slope parameter to depend on
the $p_T$ interval where the fit is carried out.

To illustrate the emission from a boosted thermal source we use the
``blast-wave'' model of Siemens and Rasmussen~\cite{Siemens} where
thermalized matter of temperature $T_f$, approximated by a boosted
Boltzmann distribution, freezes out on a thin cylindrical
shell. Assuming a boost-invariant longitudinal expansion, a transverse
flow rapidity $\rho$ on the shell and a freeze-out at constant proper
time $\tau$, the Cooper-Frye freeze-out distribution (Eq.~(\ref{CF}))
can be calculated analytically~\cite{RVesa,HLS90}. Up to irrelevant
constants one finds
\begin{equation}
  \frac{dN}{\mathrm{d}y\,m_T\,\mathrm{d}m_T} 
	\propto m_T I_0\!\left(\frac{p_T\sinh\rho}{T_f}\right)
                    K_1\!\left(\frac{m_T\cosh\rho}{T_f}\right),
  \label{blastwave}
\end{equation}
where $\rho = \tanh^{-1}v_r$ is the transverse flow rapidity.
The slope parameter $T_\mathrm{slope}$ is given by~\cite{Schnedermann:1993}
\begin{eqnarray}
  \label{blastslope}
  \frac{-1}{T_\mathrm{slope}}\,=\,\tilde{T}
     & = & \frac{\mathrm{d}}{\mathrm{d}m_T}
        \ln\left(\frac{\mathrm{d}N}{\mathrm{d}y\,m_T\,\mathrm{d}m_T}\right) \\
   & \approx & \frac{1}{m_T} + \frac{m_T}{p_T}\frac{\sinh\rho}{T}
                         - \frac{\cosh\rho}{T}, \nonumber
\end{eqnarray}
where we have approximated the Bessel functions by exponentials. The
slope parameter now has a much more complicated dependence on mass
than the linear dependence of Eq.~(\ref{Tsimple}). As seen by
differentiating Eq.~(\ref{blastslope}) with respect to mass,
\begin{equation}
  \frac{\mathrm{d}T_\mathrm{slope}}{\mathrm{d}m} 
    = \frac{1}{\tilde{T}^2}\left(-\frac{m}{m_T^3}
                                +\frac{m}{m_T p_T}\frac{\sinh\rho}{T}\right),
  \label{slopemass}
\end{equation}
there may even be a $p_T$-range where the slope parameter decreases as
a function of mass. In a more realistic calculation the emission takes
place on a surface where the flow velocity varies, and to find a $p_T$
region where the slope parameter decreases with increasing mass
requires a very large average flow velocity. At SPS and RHIC
energies the flow velocity is not sufficiently large, but the slope
parameters show a similar non-linear mass and $p_T$ dependence to that
depicted in Eq.~(\ref{blastslope}) nevertheless.

\begin{figure}
  \begin{center}
    \psfig{file=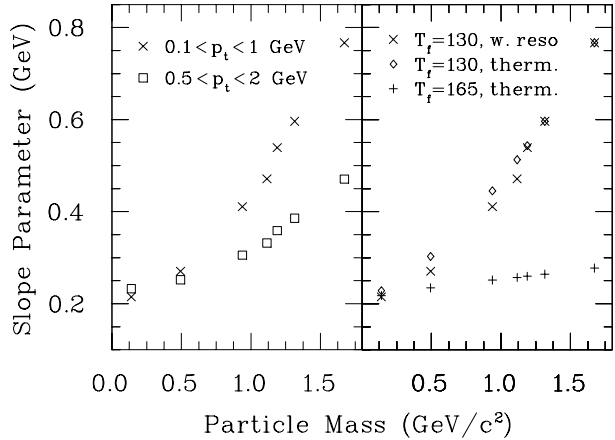,width=8cm}
  \end{center}
 \caption{Hydrodynamically calculated slope parameters as a function
          of particle mass in central collisions at RHIC 
          ($\sqrt{s}=130$ GeV/$A$). In the left panel the slope
          parameters are from fits to spectra in two different $p_T$ intervals.
          In the right panel the fits are done in interval $0.1<p_T<1$ GeV/$c$
          for spectra after resonance decays (w.\,reso), before resonance
          decays (therm) and for thermal spectra immediately after
          hadronization at $T_c=165$ MeV.} 
 \label{Tslope-fig}
\end{figure}

Besides the actual properties of boosted thermal distributions, there
is another complication in extracting flow velocity from the $p_T$
distributions. A large number of detected particles do not originate
from a thermal source but from resonance decays. Due to the available
phase space, the decays contribute mainly to the low-$p_T$ region
leading to steeper slopes than in the original thermal distribution.
In ref.~\cite{Schnedermann:1993} the slope parameter of daughter
particles originating in two body decays was approximately related to
the slope parameter of the resonance by
\begin{equation}
	T_\mathrm{eff} = \frac{p^*}{m_R}T_R,
\end{equation}
where $p^*$, $m_\mathrm{R}$ and $T_R$ are the momentum of the daughter
particle in the rest frame of the resonance, the mass of resonance and
the slope parameter of the resonance, respectively. How much the
decays change the observed slope parameter depends on the particular
resonance and particle species and the temperature where the yields of
resonances and particles are fixed. The calculation of the spectrum of
resonance products is generally a complex task and must be carried out
numerically. A detailed discussion of decay kinematics is beyond the
scope of this article and an interested reader is referred to
refs.~\cite{Schnedermann:1993,Sollfrank:1991,Hirano:2000,Florkowski:2001}
where different aspects of resonance decays have been
discussed.

How the fit intervals and resonance decays affect slope parameters can be
seen in fig.~\ref{Tslope-fig} where hydrodynamically calculated slope
parameters are shown. The calculation was tuned to reproduce the
observed $\pi$ and $p$ spectra in the most central Au+Au collisions at
$\sqrt{s}=130$ GeV/$A$ energy at RHIC~\cite{Heinz:2001}. The
slope parameters in the left panel are from fits to the same spectra
after resonance decays at freeze-out temperature at $T_f = 130$ MeV,
but the $p_T$ range where the fit was carried out was either
$0.1<p_T<1$ GeV or $0.5<p_T<2$ GeV. The pion and kaon slope
parameters turned out to be quite independent of the fit interval, but
all the heavier particles show a strong dependence on it. The slope
parameters in the right panel are all obtained using the former fit
interval, but the spectra either contained the contribution from
resonance decays (w.\,reso) or were the spectra of thermally emitted
particles (therm). For comparison the slope parameters were also
calculated for thermal spectra (T$_f$=165, therm) immediately after
hadronization at $T_c = 165$ MeV. As explained, the resonance
contribution makes the slopes steeper and decreases the slope
parameter. This effect is small for heavy particles ($m>m_{\Lambda}$)
because there are very few resonances decaying into those
particles. For the same reason, the mass dependence of the slope
parameter has a jump between the $\Lambda$ and $\Sigma$ masses.

As seen in fig.~\ref{Tslope} the experimental slope parameters of some
particles do not follow the behaviour suggested by $\pi$, $K$ and $p$.
This has been used to argue that $\phi$, $\Omega$ and $J/\Psi$ do not
experience flow in the same way as pions and protons, but decouple
earlier~\cite{NuXu}. Since the scattering cross sections of
$\phi$-mesons and $\Omega$-baryons are smaller than pions and protons,
this is possible, but as explained before, slope parameters are very
sensitive to the $p_T$ interval of the fit and thus not very reliable.
In their recent analysis~\cite{NA49:2002} the NA49 collaboration was
able to fit the $p_T$ spectra of all particles using the blast wave
model (Eq.~\ref{blastwave}) and the {\em same} parameters for all
particles including $\Omega$ and $\bar{\Omega}$. Thus the experimental
situation is unclear as to whether all hadrons experience a similar
flow or not.

Nevertheless, the almost mass independent slope parameters shown in
fig.~\ref{Tslope} do not imply absence of flow. To illustrate
this we show the slope parameters calculated from particle
distributions immediately after hadronization in
fig.~\ref{Tslope-fig}. Even if there is flow at that time, the slope
parameters show only a weak dependence on mass because the average flow
velocity is small (0.2 vs.\ 0.44 at the end of hadronic phase) and the
slope parameters are dominated by the large temperature.

  \subsection{Flow anisotropies}

The anisotropy of transverse flow is manifested as azimuthally
anisot\-ro\-pic final particle distribution. To quantify this anisotropy,
the particle spectra are expanded in harmonics of the azimuthal angle
$\phi$ event by event~\cite{Voloshin:1994}
\begin{eqnarray}
  \label{fourier}
  \frac{\mathrm{d}N}{\mathrm{d}y\mathrm{d}\phi_p} 
    & = & \frac{\mathrm{d}N}{2\pi \mathrm{d}y}(1+2v_1\cos (\phi-\phi_R)
                              +2v_2\cos 2(\phi-\phi_R) + \cdots), \nonumber \\
  \frac{\mathrm{d}N}{\mathrm{d}y \mathrm{d}p_T \mathrm{d}\phi_p}
   & = & \frac{\mathrm{d}N}{2\pi \mathrm{d}y \mathrm{d}p_T}
                            (1+2v_1(p_T)\cos (\phi-\phi_R)\\
   &   & \qquad\qquad\quad +{}2v_2(p_T)\cos 2(\phi-\phi_R) + \cdots), \nonumber
\end{eqnarray}
where $\phi_R$ is the azimuthal angle of the reaction plane. Assuming
that the experimental uncertainties in event plane reconstruction can
be corrected for, each event can be rotated such that $\phi_R=0$. The
expansion parameters $v_1$ and $v_2$ correspond to \emph{directed} and
\emph{elliptic} flow, respectively. Due to symmetry both vanish in central
collisions which are cylindrically symmetric. In symmetric collision
systems the odd coefficients $v_1, v_3,\ldots,$ are zero at
midrapidity.

\begin{figure}[t]
  \begin{center}
     \psfig{file=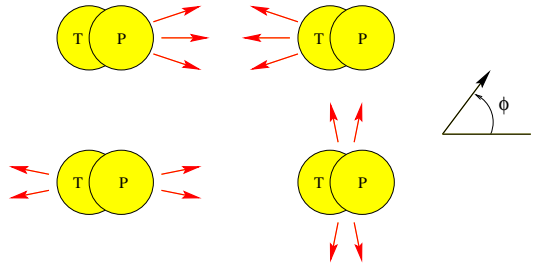,width=7cm}
  \end{center}
  \caption{Schematic representation of the collision geometry and
           different anisotropies of flow seen in the transverse plane.
           P and T denote the projectile and target nuclei, respectively.
           Top: directed flow in projectile rapidity region, positive (left)
           and negative (right). Bottom: elliptic flow, in plane (left) and
           out of plane (right). From ref.\protect\cite{Ollitrault:1997}.}
  \label{skema}
\end{figure}

  \subsubsection{Directed flow}
     \label{v1}

At AGS energies the general picture of the origin of directed
flow~\cite{Herrmann} is that the pressure formed in the collision
region deflects the projectile and intermediate rapidity fragments,
i.e.\ spectator nucleons away from the target (``bounce-off'' and
``sidesplash'' effects~\cite{ur-flow}) resulting in a preferred
direction of nucleon emission. Simultaneously the produced pions
scatter from spectator nucleons forming resonances. Through this
process the initial direction of pions is lost and flow into the
direction of spectator matter is reduced. Thus pions show flow to the
opposite direction than nucleons. The usual sign convention is to
choose protons to have positive directed flow in the projectile
rapidity region and negative in the target rapidity region. Directed
flow is zero at midrapidity due to symmetry and saturates before
projectile and target rapidities are reached forming an overall
s-shaped curve as function of rapidity~\cite{Herrmann}.

The strength of directed flow depends on the pressure formed in the
collision region and the time that the secondary particles have to
interact with the spectators, i.e.\ how fast the spectators pass by
the collision zone. This time gets smaller with increasing energy and
one expects the directed flow at SPS be smaller than at AGS, as is
observed~\cite{Herrmann}. The formation of directed flow in the very
early stages of the collision in very short times poses a problem for
the hydrodynamical calculation. Especially at SPS and RHIC
energies it is possible that most of directed flow is established
before the system has reached local thermal equilibrium and the
pre-equilibrium features dominate~\cite{Sorge:1996}.

So far there are no detailed hydrodynamic calculations of directed
flow at SPS or RHIC energies. However, it has been predicted that a
phase transition in a nuclear equation of state would lead to a local
minimum of directed flow as a function of collision
energy~\cite{Rischke:1995}. The exact position and value of this
minimum depend on the details of the model, but it should be located
between the AGS and the maximum SPS energy.

Another interesting prediction is that when the collision energy
increases, the rapidity dependence of directed flow would eventually
have such a shape that the sign of directed flow changes three times
as a function of rapidity~\cite{Brachmann:1999,Csernai:1999,Snellings:1999}.
In refs.~\cite{Brachmann:1999} and~\cite{Csernai:1999} this behaviour
is called ``antiflow'' and ``third flow component'', respectively, and
its origin is explained by the phase transition and the initial shape
of the system. The phase transition leads to an initial shape of a
disk slightly tilted away from the beam axis. Thus the largest
pressure gradient and the direction of fastest expansion points away
from the spectator nucleons and the direction of the conventional
directed flow. Close to midrapidity the emission from the collision
region dominates and the particles show negative directed flow in the
forward and positive in the backward rapidity region, whereas close to
fragmentation regions directed flow is again due to interactions with
spectator matter and has its usual sign. The initial argument of
ref.~\cite{Csernai:1999} was for AGS energies, but has been refined
for RHIC~\cite{Magas:2000}.

Similar behaviour for a completely different reason is predicted in
ref.~\cite{Snellings:1999} where the changes of sign are predicted to
happen above SPS energies independent of a phase transition. Their
argument is that in non-central collisions nuclear transparency leads
to an anisotropic distribution of baryon number in the initial state,
which is then manifested as anisotropic particle emission: when the
particle fluid expands, baryons are carried to the same side as they
were at the beginning of expansion. Even if the authors do not mention
it, these two approaches can be expected to have observable
differences in pion directed flow: emission from a tilted
disk~\cite{Magas:2000} is similar for both pions and protons and both
should show directed flow in the same direction close to midrapidity
and in opposite directions in fragmentation regions.  On the other
hand, an inhomogeneous source~\cite{Snellings:1999} leads to pion flow
in the opposite direction to proton flow at all rapidities, or
alternatively to a negligibly small directed flow of pions. From the
hydrodynamical point of view, both of these approaches are possible
which highlights the uncertainties in choosing the initial state of
hydrodynamic evolution (section~\ref{init}).

Unfortunately the SPS data~\cite{NA49,WA98-v1} does not have a large
enough rapidity coverage to test these ideas. So far there is no
measurement of directed flow at RHIC. Whether it is large enough to be
experimentally observable remains to be seen.

  \subsubsection{Elliptic flow}
     \label{v2}

\begin{figure}[t]
\epsfxsize=8.6truecm
 \begin{center}
   \mbox{\psfig{file=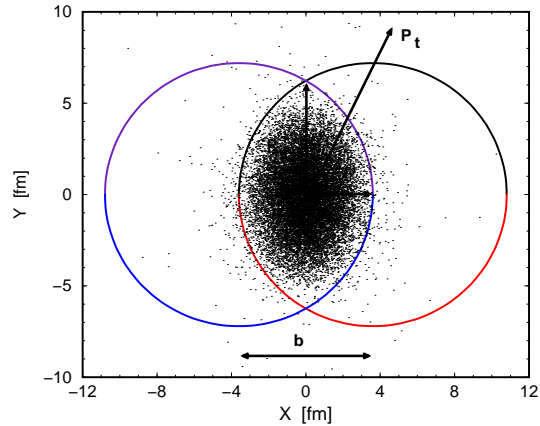,height=8.cm,angle=-90}}
 \end{center}
    \caption{Reaction plane of a semi-central $Au+Au$ collision for impact
             parameter $b = 7$ fm. The density of dots is proportional to
             the number of participating nucleons in the overlap region.
             From ref.\protect\cite{hh}.}
    \label{origin}
\end{figure}

In non-central collisions the primary particle production is
azimuthally isotropic, but since the interaction region is
aniso\-tropic in space, the secondary collisions can cause the final
particle distribution to be anisot\-ro\-pic in momentum space. As
depicted in fig.~\ref{origin}, particles heading out of plane have on
the average a longer distance to go within the dense region than
particles moving in plane. Thus particles moving out of plane have a
larger probability to scatter several times and change their direction
than particles heading in plane. Thus the final particle distribution
has more particles moving in the in plane than the out of plane
direction and the Fourier coefficient $v_2$ in Eq.~\ref{fourier} is
positive.

In a hydrodynamic picture the buildup of such elliptic anisotropy can
be understood in terms of pressure gradients. The average pressure
gradient between the center of the system and the surrounding vacuum
is larger in plane than out of plane direction because the system is
thinner in that direction. Consequently the collective flow velocity
is larger in plane than out of plane. This leads to a larger average
momentum in plane than out of plane and to more particles being
emitted in the in plane than the out of plane direction.

The origin of elliptic anisotropy is thus rescatterings and the shape
of the system. The asymmetry of the shape of the system is largest
immediately after the collision and decreases with increasing time
independently of the frequency of rescatterings. Only the speed with
which the system will gain an azimuthally symmetric shape depends on
how frequent the rescatterings are. Thus a large elliptic anisotropy
and a large value of $v_2$ is taken to be a signal of abundant
rescatterings in the early stage of the collision and thus a signal of
early pressure buildup and thermalization~\cite{Sorge:1996,Heinz:2001}.
Since hydrodynamics assumes zero mean free path and thus infinite
rescattering, it is also assumed to give the practical upper limit of
elliptic anisotropy.

The relation between large $v_2$ and early pressure buildup is
difficult to quantify because of the uncertainty of the initial time
of the hydrodynamical evolution. If pressure buildup and
thermalization take a ``long'' time, say several fm/c, the particles
formed in the primary collisions can move significantly in the
transverse plane. The geometric arguments presented in
section~\ref{init} to constrain the initial shape are no longer valid
and the initial state of hydrodynamical evolution is largely
unknown. Consequently it is not possible to calculate reliably the
elliptic flow parameter $v_2$ as function of thermalization time
$\tau_0$. The uncertainty of shape also explains the conflicting
results in the literature: if the shape does not change, $v_2$ is
almost independent of $\tau_0$
\cite{Ollitrault:1992}.  On the other hand, if the shape changes as if
the particles were freely streaming and the ratio of the anisotropy of
the initial shape and the final $v_2$ stays the same, a delay of 3.5
fm/c in thermalization was estimated to reduce $v_2$ by
50\%~\cite{Kolb:2000a}.

\begin{figure}
  \begin{center}
    \psfig{file=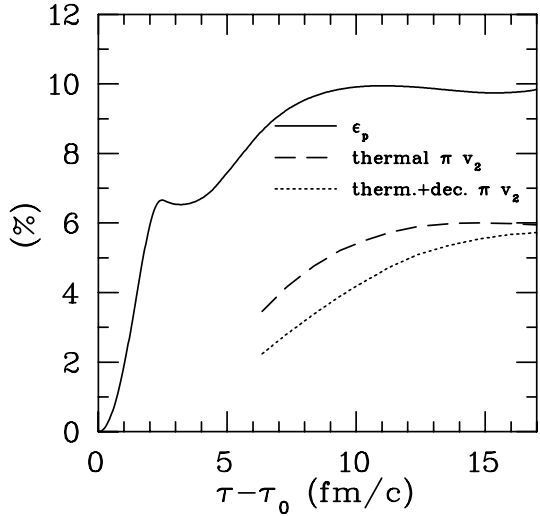,width=7cm}
  \end{center}
    \caption{Time-evolution of momentum anisotropy (solid line),
             elliptic anisotropy of thermal pions (dashed line) and
             a sum of thermal pions and pions originating from resonance
             decays (dotted line) in Au+Au collision at
             $\sqrt{s}=130$ GeV/A with impact parameter $b=7$ fm.}
    \label{v2evo}
\end{figure}

To gain insight how elliptic anisotropy is built up during the
hydrodynamical evolution one uses so called momentum
anisotropy~\cite{Kolb:2000a}, labeled either $\delta$ or
$\epsilon_x$ in the literature,
\begin{equation}
  \epsilon_x = \delta 
    = \frac{\langle T^{xx}-T^{yy}\rangle}{\langle T^{xx}+T^{yy}\rangle}
    = \frac{\int \mathrm{d}x\,\mathrm{d}y\,(T^{xx}-T^{yy})}
           {\int \mathrm{d}x\,\mathrm{d}y\,(T^{xx}+T^{yy})},
\end{equation}
where $T^{xx}$ and $T^{yy}$ are the components of energy-momentum
tensor $T^{\mu\nu}$, the angle brackets denote averaging over the
transverse plane and integration is done over transverse plane at
constant proper time $\tau$. As an example the time evolution of
momentum anisotropy (solid line) in a simulated Au+Au collision at
$\sqrt{s} = 130$ GeV/A collision energy is shown in fig.~\ref{v2evo}.

The shoulder in the increase of momentum anisotropy can be related to
the phase transition~\cite{Kolb:2000a}. In this particular case it
occurs when no part of the system is in the plasma phase anymore, but
the anisotropy begins to increase before the system is entirely
hadronized.  As shown in ref.~\cite{Kolb:2000a}, the actual effect of
a phase transition on the buildup of momentum anisotropy is
complicated and depends on the details of the flow field at the time
of the phase transition and the relative sizes of the plasma, mixed
and hadronic phases at each point of time. It is possible that the
increase of anisotropy is halted as in fig.~\ref{v2evo}, but it is
equally possible that anisotropy keeps increasing or even decreases
during the mixed phase. The decrease during the mixed phase can be
explained by noting that the pressure gradients within the mixed phase
are tiny and therefore the matter in the mixed phase keeps flowing
with the flow velocity it had when entering the mixed phase. When the
matter flows outwards with constant velocity, its flow is nearly
self-similar, which decreases the momentum anisotropy. Whether this
effect can be seen in the momentum anisotropy of the \emph{entire}
system, depends on how large a part of the system is in mixed phase.

Another feature of the time evolution of the momentum an\-isot\-ro\-py
is that after saturating at $\tau-\tau_0 \approx 10$ fm, it begins to
slightly decrease. This is a result of the self-cancelling nature of
the anisotropy; the anisotropic pressure gradients cause the system to
expand more strongly in-plane than out of plane. In the course of time
the system shape becomes cylindrically isotropic, but due to momentum
conservation the larger expansion in-plane continues.  The system
becomes elongated \emph{in-plane} and the anisotropy of the pressure
gradient begins to act against the momentum anisotropy.

The relationship between the momentum anisotropy, $\epsilon_p$, and the
observed elliptic anisotropy, $v_2$, is nontrivial. The former measures
the anisotropy of the collective flow velocity, whereas the latter the
an\-isot\-ro\-py of particle yield. The particle distributions are a result
of both collective and thermal motion and therefore the ratio of
$\epsilon_p$ and $v_2$ depends on the mass of the particle in question
and the freeze-out temperature. To illustrate this dependence we show
in fig.~\ref{v2evo} also $v_2$ for pions calculated assuming
freeze-out at various temperatures $165<T_f<90$ MeV and plotted as
function of the freeze-out time of the system (dashed line). It can be
seen that the momentum anisotropy increases roughly by 25\% after the
system is completely hadronized (at $\tau-\tau_0\approx 6.5$ fm/c),
whereas the anisotropy of pions increases by 70\% -- the difference
caused by the increasing sensitivity of particle distributions to
collective motion when temperature decreases.

Another complication in comparing the momentum anisotropy of the flow
to the observed anisotropy of particle distributions are the resonance
decays. The kinematics of decay not only favours the daughter
particles having a smaller $p_T$ than the decaying resonance, but also
an azimuthal distribution of daughter particles which is peaked
out-of-plane even if the distribution of the resonances peaks
in-plane~\cite{Hirano:2000}.  This leads to daughter particles showing
\emph{negative} elliptic anisotropy, which dilutes the anisotropy of
particles with thermal origin, as can be seen in fig.~\ref{v2evo},
where the dotted curve shows the $v_2$ of pions originating both from
decays and a thermal source. The effect of resonance decays depends on
temperature because the relative abundances of particles and
resonances depend on temperature -- the higher the temperature, the
larger the fraction of pions that originates from resonance decays.

\begin{figure}[t]
  \begin{center}
    \psfig{file=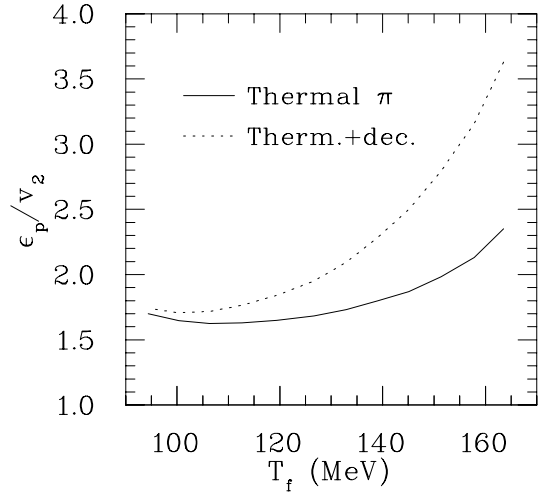,width=7cm}
  \end{center}
    \caption{The ratio of momentum anisotropy, $\epsilon_p$, to elliptic
             anisotropy, $v_2$, of pions as function of freeze out temperature
             $T_f$ in Au+Au collisions at $\sqrt{s}=130$ GeV/A with impact
             parameter $b=7$ fm.}
    \label{ratio}
\end{figure}

We also show the ratio $\epsilon_p/v_2$ as function of temperature in
fig.~\ref{ratio}. It is seen that $v_2$ for thermal pions is
roughly half of the momentum anisotropy, but when resonance decays are
included, the ratio is much more temperature dependent.

\begin{figure}[t]
  \begin{minipage}{5.5cm}
  \begin{center}
    \psfig{file=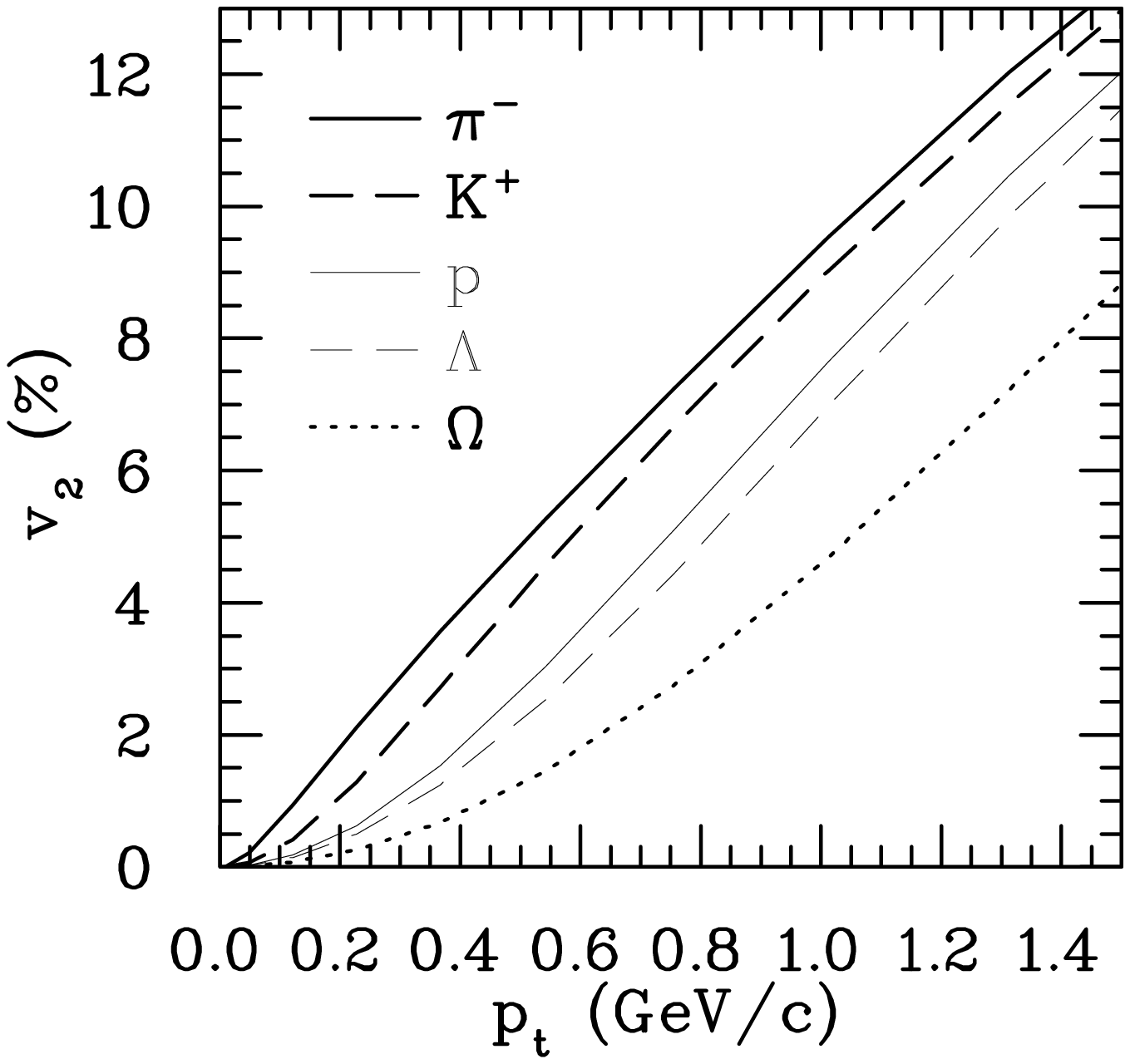,width=5.5cm}
  \end{center}
  \end{minipage}
     \hfill
  \begin{minipage}{5.5cm}
    \psfig{file=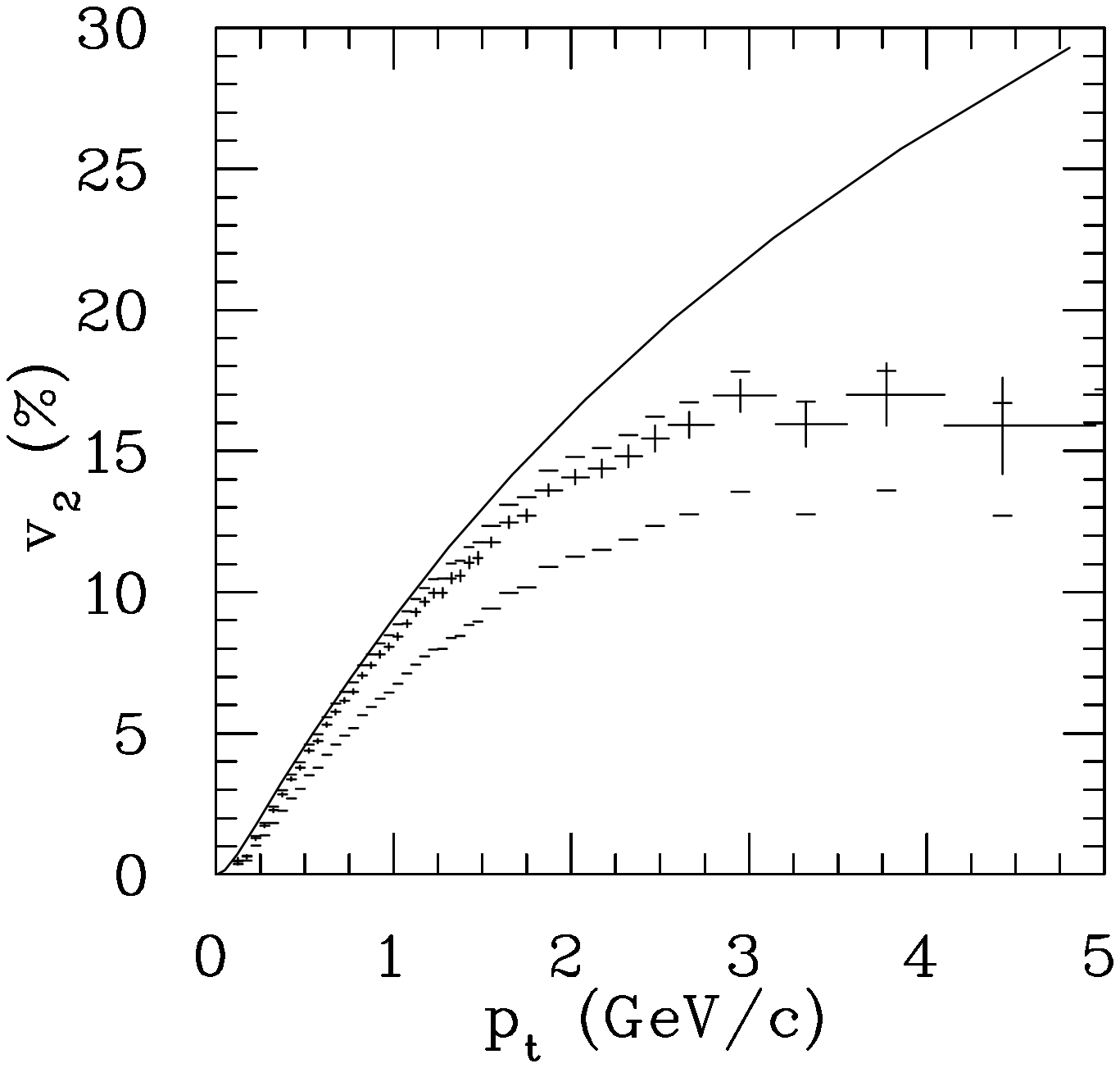,width=5.5cm}
  \end{minipage}
    \caption{The calculated $p_T$ differential elliptic anisotropy
             coefficient $v_2(p_T)$ at midrapidity in minimum bias
             Au+Au collisions at $\sqrt{s}=130$ GeV/A. The left figure
             shows $v_2(p_T)$ for various particle
             species\protect\cite{Huovinen:2001} and the right figure a
             comparison of the calculated $v_2(p_T)$ for charged
             particles\protect\cite{Heinz:2001} with STAR
             data\protect\cite{STAR-highpt}.}
    \label{v2pt}
\end{figure}

In general the heavier the particle, the more sensitive it is to
collective motion. For example, as explained in chapter \ref{pt}, if
the freeze-out temperature and flow velocity are the same, the heavy
particles have flatter $p_T$ distributions and thus larger slope
parameters $T_\mathrm{slope}$ than lighter particles. In the same way,
the $p_T$ averaged elliptic anisotropy parameter $v_2$ is observed to
increase with particle mass. However, at low values of $p_T$, the
$p_T$-differential anisotropy parameter $v_2(p_T)$, shows the opposite
behaviour: the heavier the particle, the smaller the value of
$v_2(p_T)$ at fixed $p_T$ (see fig.~\ref{v2pt}).

The apparent contradiction between $v_2$ and $v_2(p_T)$ has a simple
explanation. $v_2$ is not an additive quantity, but when a
$p_T$-averaged $v_2$ is calculated from $v_2(p_T)$,
the latter is weighted by the particle distribution:
\begin{equation}
  v_2 = \frac{\int \mathrm{d}p_T\,v_2(p_T) \frac{\mathrm{d}N}{\mathrm{d}p_T}}
                {\int \mathrm{d}p_T\, \frac{\mathrm{d}N}{\mathrm{d}p_T}}.
\end{equation}
Thus the flatter $p_T$ distribution of heavier particles weights the
high $p_T$ region, where $v_2(p_T)$ is larger, more and the $p_T$-averaged
value can be larger. Whether this larger weight for high $p_T$ wins over
the reduction of $v_2$ at fixed $p_T$ depends on the details of the
expansion dynamics and the contribution from resonance decays.

The mass dependence of the elliptic anisotropy at fixed $p_T$ can be
explained as an interplay between transverse collective flow, random
thermal motion and anisotropy of the flow field~\cite{Huovinen:2001}.
It is well known that transverse flow shifts the $p_T$-distributions
to larger values of $p_T$. For nonrelativistic $p_T < m$ this effect
increases with the particle mass $m$ and the transverse flow velocity
$\langle v_\perp \rangle$. In the extreme case of a thin shell
expanding at high velocity, the spectrum actually develops a relative
minimum at $p_T=0$ and a peak at nonzero $p_T$ (``blast wave peak''
\cite{Siemens}), and with increasing mass the peak shifts to larger
$p_T$. Relative to the case without transverse flow, the spectrum is
thus depleted at small $p_T$, and the depletion, as well as the $p_T$
range over which it occurs, increase with $m$ and
$\langle v_\perp \rangle$.

If the transverse velocity is larger in the $x$ than the $y$
direction, the same is true for this relative depletion effect. It
counteracts the overall excess of particles moving to the $x$
direction over particles moving to the $y$ direction reducing
$v_2$. This reduction and the range where it occurs increases with
particle mass and average transverse flow velocity $\langle v_\perp
\rangle$. In the extreme case of a thin shell, this depletion effect
can be so large that there are less small $p_T$ particles moving in
the $x$ than the $y$ direction, and $v_2$ becomes negative at low
$p_T$. When a constant expansion velocity of a thin shell is replaced
by a realistic transverse velocity distribution, the peak in the
single particle spectrum disappears~\cite{Lee:1990}. In the same way
a realistic velocity profile weakens the reduction of $v_2$ at low
$p_T$, but the mass dependence of $v_2$ at low $p_T$ remains. Whether
particles show a positive or negative $v_2$ at low $p_T$ depends on
the details of the flow profile.

For relativistic $p_T > m$, the particle mass does not play a role in
the thermal distribution, and consequently the curves showing
$v_2(p_T)$ approach each other. In a simple model where the transverse
velocity profile is replaced by its average value, as in the blast
wave model of Siemens and Rasmussen~\cite{Siemens} and its
derivatives~\cite{Huovinen:2001,Adler:2001}, $v_2$ increases with
$p_T$ and approaches an asymptotic value of one. The details of the
velocity profile can change this behaviour, but so far
hydrodynamical calculations with realistic initial conditions have
shown similar monotonic increases of $v_2$ with $p_T$ (see
fig.~\ref{v2pt}).

\section{Learning from RHIC data}
  \label{data}

\begin{figure}
  \begin{center}
    \psfig{file=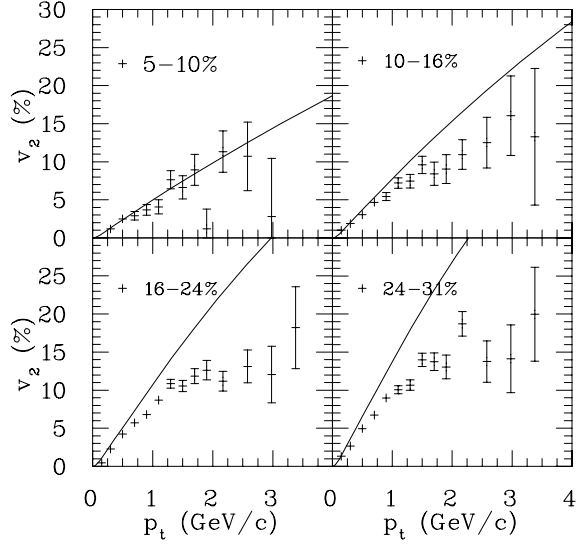,width=7.5cm}
  \end{center}
    \caption{The differential elliptic anisotropy $v_2(p_T)$ of charged
             particles in $Au+Au$ collisions at $\sqrt{s}=130$
             GeV/A at various centralities compared with the STAR
             data\protect\cite{4-ple}. The calculation is similar to that of
             ref.\protect\cite{Heinz:2001}.}
    \label{binbybin}
\end{figure}

\begin{figure}[t]
  \begin{center}
    \psfig{file=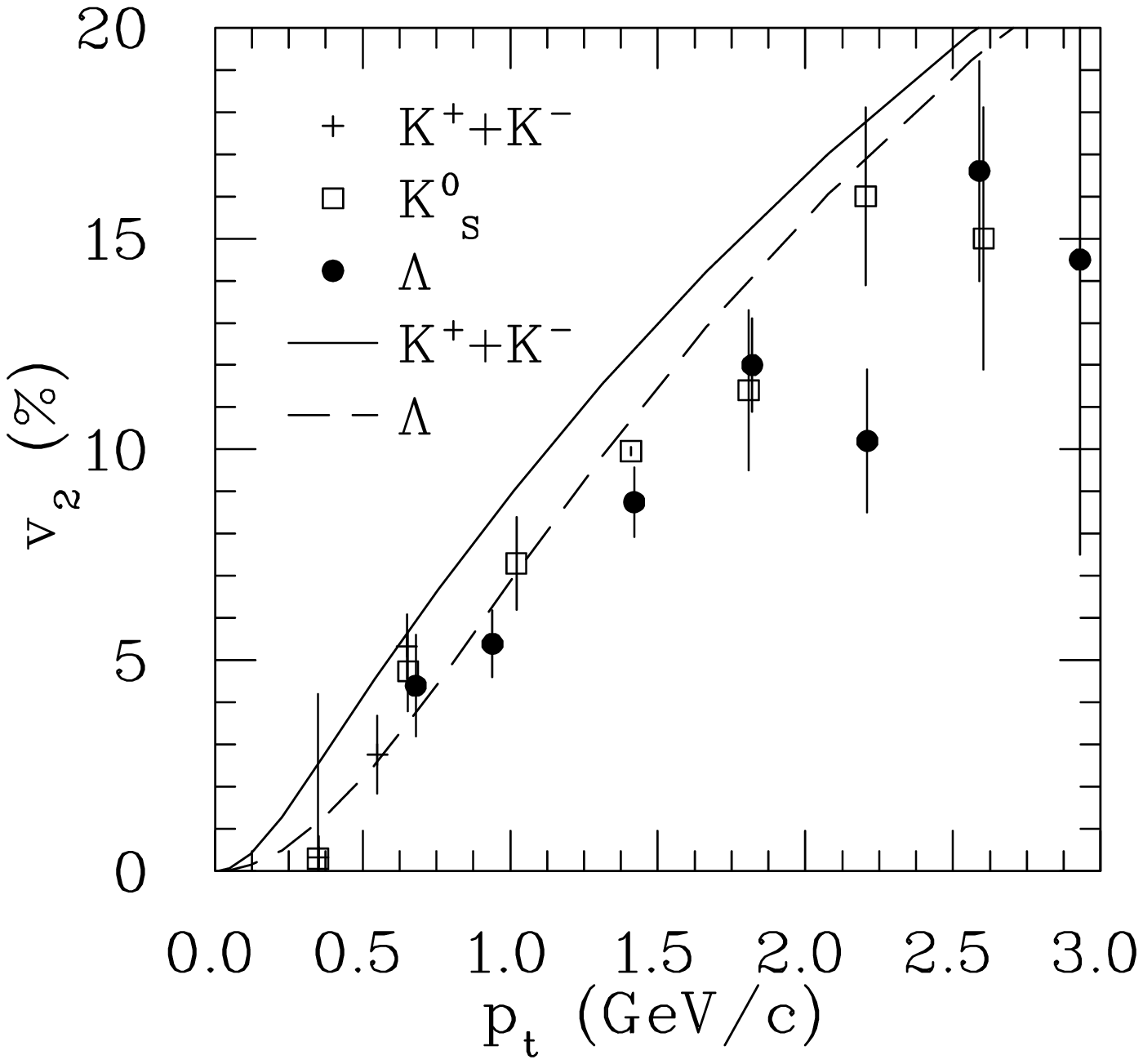,width=6.5cm}
  \end{center}
    \caption{The differential elliptic anisotropy $v_2(p_T)$ of kaons and
             Lambdas in minimum bias $Au+Au$ collisions at $\sqrt{s}=130$
             GeV/A compared with the STAR data\protect\cite{Adler:2001,kaonv2}.
             The calculation is similar to that of
             ref.\protect\cite{Heinz:2001}. }
    \label{strange}
\end{figure}

\begin{figure}
  \begin{center}
    \psfig{file=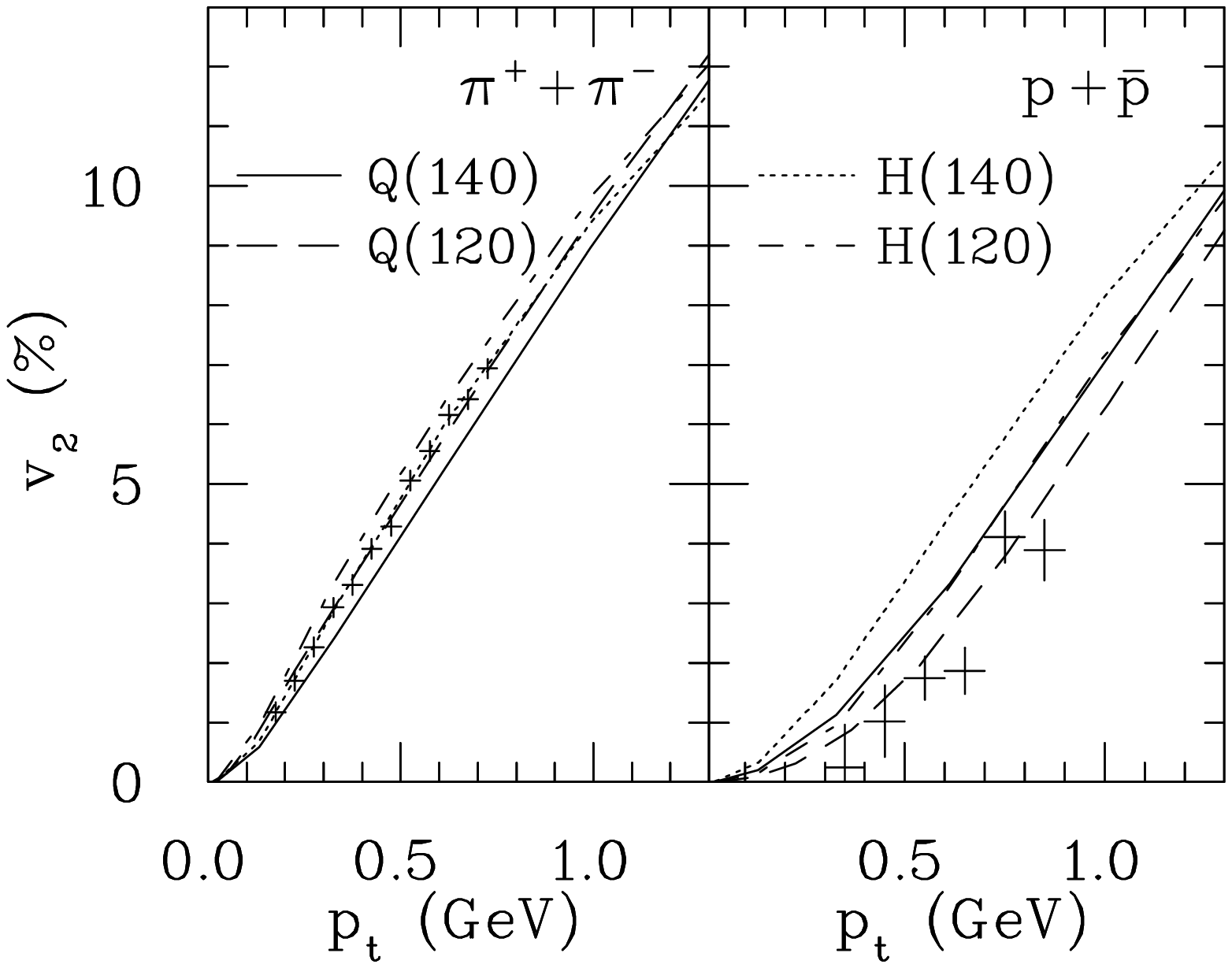,width=7.5cm}
  \end{center}
    \caption{The differential elliptic anisotropy $v_2(p_T)$ of
             pions and protons in minimum bias $Au+Au$ collisions
             at $\sqrt{s}=130$ GeV/A\protect\cite{Huovinen:2001} compared
             with the STAR data\protect\cite{Adler:2001}. The letters Q
             and H in the labels stand for an equation of state with a first
             order phase transition and a hadron gas equation of state without
             a phase transition, respectively. Numbers in parentheses
             stand for the freeze-out temperature in MeV.}
    \label{eos}
\end{figure}

One of the first measurements of $Au+Au$ reactions at the Relativistic
Heavy Ion Collider (RHIC) at Brookhaven National Laboratory (BNL) at
$\sqrt{s}=130$ GeV/$A$ energy was a measurement of elliptic flow at
midrapidity~\cite{STAR-v2}. Soon afterwards the elliptic flow in
minimum bias collisions was shown to be very nicely reproduced by a
hydrodynamical model~\cite{Kolb:2000b}. As shown in fig.~\ref{v2pt},
the differential anisotropy $v_2(p_T)$ in minimum bias collisions
follows the hydrodynamical curve closely up to momentum
$p_T = 1$--1.5\,GeV As a function of centrality the hydrodynamical
description works as well or even better up to 16\% of the most central
collisions ($b\ltsim 6$ fm), see fig.~\ref{binbybin}. At larger impact
parameters, the $p_T$ region where the hydrodynamical calculation fits
the data becomes smaller and the deviation from the hydrodynamical curve
grows faster. The identified particle anisotropy shows a similar mass
ordering to that discussed at the end of previous section
(figs.~\ref{strange} and~\ref{eos}), although the quantitative fit,
especially of kaons, is not as good as the fit of pions and charged
particles. The system thus behaves as a thermal system and this result
has been taken as an indicator of thermalization~\cite{Heinz:2001}.

Since flow is closely connected to the equation of state of matter, it
is possible to use details of the flow to constrain it. However, flow is
equally sensitive to freeze-out temperature and the initial state of the
evolution, so to learn about equation of state one must also know
how flow depends on all other variables.

The basic rule as to how $p_T$ averaged anisotropy depends on the
freeze-out temperature at RHIC energies is simple. The smaller the
freeze-out temperature, the larger the $v_2$ --- if everything else in
the simulation stays unchanged. This rule is not valid
indefinitely. As explained in Section~\ref{v2} and seen fig.~\ref{v2evo},
when the shape of the system has become azimuthally symmetric, the
pressure gradients begin to work against the buildup of anisotropy and
the momentum anisotropy begins to decrease. The observed $v_2$ may
still keep increasing after the momentum anisotropy has begun to
decrease because lower temperature makes the final particle
distributions more sensitive to the anisotropy of collective motion,
but eventually it will begin to decrease too.

Unfortunately the temperature dependence of the $p_T$ differential
anisotropy, $v_2(p_T)$, is much more complicated. It is different for
each particle species and it is also sensitive to the equation of
state used. As shown in Fig.~\ref{eos}, a decrease of freeze-out
temperature from 140 to 120 MeV, increases the pion $v_2(p_T)$
slightly, but decreases the proton $v_2(p_T)$. This again is a result
of an interplay of collective flow and random thermal motion which is
different for particles of different mass.  The change in freeze-out
temperature means two things: first, the lifetime of the system
changes and therefore the amount of flow changes. Second the random
thermal motion changes. Since the observed particle distributions are
the result of both, the changes in $v_2$ result from both. Intuitively
the effect of freeze-out temperature can be explained using the blast
wave model in the same way as the mass dependence of $v_2(p_T)$
(Section \ref{v2})). When freeze-out temperature decreases, transverse
flow velocity increases and the blast wave peak in the proton $p_T$
distribution becomes more prominent and moves to larger $p_T$.  This
makes the relative depletion effect at low $p_T$ larger and thus
decreases the anisotropy. The pion $p_T$ distribution on the other
hand does not show a similar peak or it lies at very small values of
$p_T$. Therefore there is no such depletion effect which would
decrease $v_2$ or it is limited to very low values of $p_T$.

\begin{figure}[t]
  \begin{center}
    \psfig{file=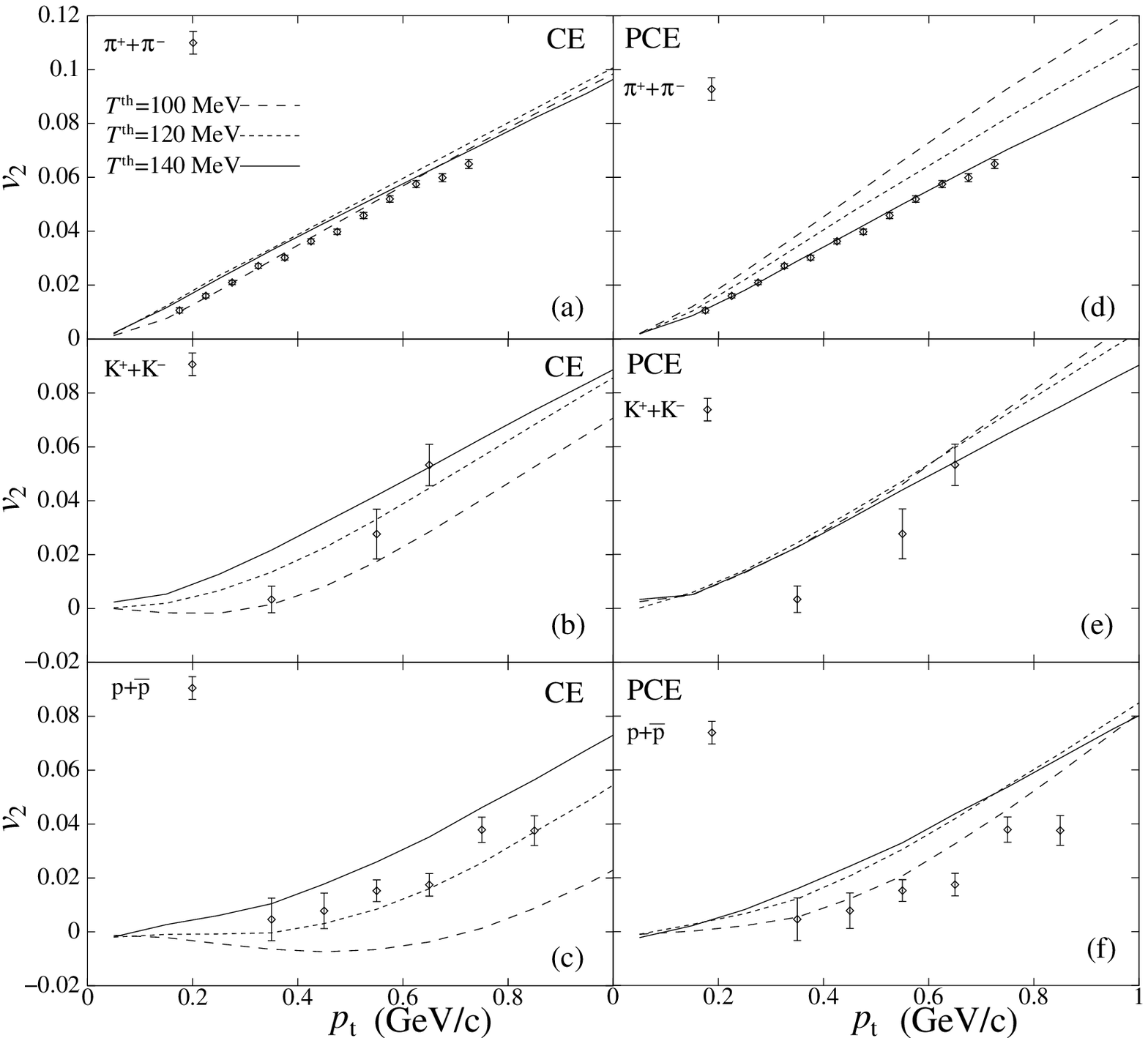,width=7cm}
  \end{center}
    \caption{The differential elliptic anisotropy $v_2(p_T)$ of pions, kaons
             and protons in minimum bias $Au+Au$ collisions at $\sqrt{s}=130$
             GeV/A at different kinetic freeze-out
             temperatures\protect\cite{Hirano:2002}
             compared with the STAR data\protect\cite{Adler:2001}.
             The left and right panels represent the results calculated
             assuming local chemical equilibrium until kinetic
             freeze-out (CE) or chemical freeze-out already at
             $T_{ch}=170$ MeV (PCE).}
    \label{chem}
\end{figure}

Fig.~\ref{eos} also shows $v_2(p_T)$ calculated using equations of
state (EoS) with a phase transition (Q) and without a phase transition
(H). In both cases the change in freeze-out temperature changes
$v_2(p_T)$ as described. This is not surprising since below the phase
transition temperature $T_c=165$ MeV of EoS Q, both equations of state
are identical. On the other hand, if the hadronic part of the equation
of state is different, a change in the freeze-out temperature may
affect $v_2(p_T)$ differently. This is shown in fig.~\ref{chem} where
$v_2(p_T)$ for pions, kaons and protons is calculated assuming either
local chemical equilibrium until kinetic freeze-out (CE), or that the
relative abundancies of each particle species was frozen out
immediately after phase transition at $T_{ch}=170$ MeV (PCE). In other
words, the chemical freeze-out takes place at much larger temperature
than kinetic freeze-out.

In the case of chemical equilibrium (CE) the equation of state is very
similar to EoS Q used to calculate the results shown in fig.~\ref{eos}
and the freeze-out temperature dependence is similar. The only
difference is that when temperature decreases from 120 MeV to 100 MeV,
the pion $v_2(p_T)$ does not increase anymore but begins to decrease.
If chemical equilibrium is lost already at hadronization, the
behaviour changes. The proton $v_2(p_T)$ still decreases with
temperature, but the dependence is much weaker. On the other hand the
pion $v_2(p_T)$ increases when $T_f$ decreases in the entire
temperature interval $100 < T_f < 140$ MeV and the increase is much
stronger than in the case of chemical equilibrium. The most dramatic
change is in the kaon $v_2(p_T)$: if chemical equilibrium is
maintained, $v_2(p_T)$ decreases when $T_f$ decreases, but in the case
of non-equilibrium, the behaviour is opposite. The kaon $v_2(p_T)$
behaves like that of pions and increases as temperature decreases.
Thus there are no simple rules to tell how the observed $v_2(p_T)$
would correspond to a certain freeze-out temperature, but the
constraints must be searched case by case for each equation of state.

In general a stiffer equation of state causes larger flow. This holds
also for elliptic anisotropy and, at least in the case shown in
fig.~\ref{eos}, also for $p_T$ differential anisotropy: EoS H leads to
a larger $v_2(p_T)$ for both pions and protons than EoS Q. For the
chemical equilibrium or out-of-equilibrium equations of state in
fig.~\ref{chem} this rule is less clear because both equations of
state are almost equally stiff --- the pressure as a function of
energy density is almost identical in both cases. The difference
between CE and PCE comes from the relation between temperature and
energy density. The same energy density corresponds to a smaller
temperature in PCE than in CE and therefore the system cools
faster. Correspondingly the flow at fixed temperature is smaller when
the system is out of chemical equilibrium and the final anisotropy
looks very different in these two cases. Change to chemical
non-equilibrium means ending the evolution at an earlier stage, but
the effect is not as straightforward as increasing the freeze-out
temperature, because the random thermal motion is not changed.

Constraining the equation of state by the transverse momentum spectra
alone is notoriously difficult since larger flow generated by a
stiffer equation of state can to a large extent be compensated by a
higher freeze-out temperature and a slightly different initial
state~\cite{Huovinen:1998}. As can be seen in fig.~\ref{eos},
essentially the same holds for pion elliptic flow: a purely hadronic
equation of state (H) and an equation of state with a phase transition
(Q) create similar $v_2(p_T)$ for pions if the freeze-out temperature
is chosen to be $T_f=140$ and 120 MeV for EoS H and Q,
respectively. On the other hand, the effect on the proton $v_2(p_T)$
is exactly opposite: this choice leads to the largest difference of
all the combinations studied. The different sensitivity of the proton
and pion $v_2(p_T)$ to the equation of state and freeze-out
temperature gives an additional handle on constraining both.

The results in fig.~\ref{eos} clearly favour an equation of state with
a phase transition. As well one could claim, based on fig.~\ref{chem}
that the proton $v_2(p_T)$ favours an equation of state where chemical
equilibrium is maintained for temperatures lower than to $T_{ch}=170$
MeV. Unfortunately both conclusions are premature since there are
other variables to be considered which affect the observed
differential anisotropy.

\begin{figure}[t]
  \begin{center}
   \psfig{file=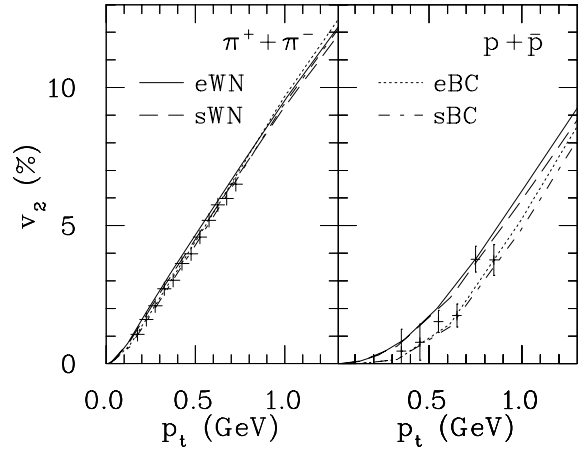,width=7.5cm}
  \end{center}
    \caption{The differential elliptic anisotropy $v_2(p_T)$ of
             pions and protons in minimum bias $Au+Au$ collisions
             for different initialization models\protect\cite{Kolb:2001b}
             compared with the STAR data\protect\cite{Adler:2001}.}
    \label{initialization}
\end{figure}

As explained in Section~\ref{init}, there are various ways to
parametrize the initial state of hydrodynamical evolution. The
simplest ones described in Section~\ref{init} all lead to slightly
different shapes of the initial system. If the freeze-out temperature
is kept unchanged, the ratio of the initial spatial anisotropy and the
observed elliptic anisotropy is almost independent of impact parameter
in a hydrodynamical model~\cite{Ollitrault:1992,Kolb:2000a}. Despite
that, the initial shape has its effect on the differential anisotropy
$v_2(p_T)$ in minimum bias collisions. The anisotropy calculated using
the initial state parametrizations described in Section~\ref{init} are
shown in fig.~\ref{initialization} \cite{Kolb:2001b}. The pion
anisotropy shown in the left panel is practically independent of
initialization, but the proton $v_2(p_T)$ is clearly sensitive to the
initial shape. These results shown in fig.~\ref{eos} were obtained
using eWN initialization and in principle it is possible that
combining sBC initialization with EoS H would the reduce proton
anisotropy sufficiently to reach the data.

The most comprehensive study of $p_T$ spectra and anisotropies at both
SPS and RHIC so far~\cite{Teaney:2001} reached the conclusion that an
equation of state with a phase transition is necessary for a
consistent reproduction of the data. This study used a hybrid model of
hydro and transport where the system evolved hydrodynamically until
hadronization and the hadron phase was described using a RQMD
transport model. In this way the uncertainty of chemical
non-equilibrium and its effects was circumvented. The freedom in
initial parametrization, however, was not explored, only the
initialization sWN being used. Thus in the framework of a
hydrodynamical description it is safe to say so far that the data
seems to favour an equation of state with a phase transition, but no
final conclusion can be drawn yet.

\begin{figure}[t]
  \begin{center}
    \psfig{file=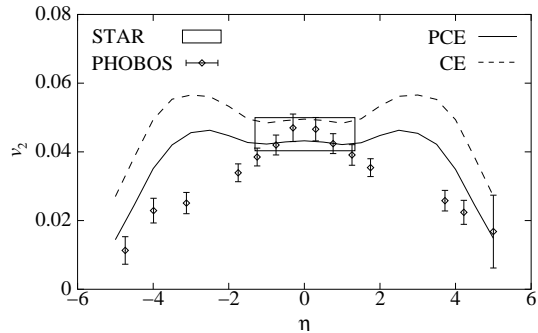,width=7cm}
  \end{center}
   \caption{Pseudorapidity dependence of elliptic anisotropy in minimum
            bias Au+Au collisions at $\sqrt{s}=130$ GeV/A compared to
            STAR\protect\cite{STAR-v2} and PHOBOS\protect\cite{phobos}
            data assuming chemical freeze out at kinetic freeze out
            (CE) or at hadronization (PCE)\protect\cite{Hirano:2002}.}
  \label{eta}
\end{figure}

From the hydrodynamical point of view elliptic flow at forward and
backward rapidities is governed by the very same physics that creates
elliptic flow at mid-rapidity. Thus to reproduce elliptic flow as a
function of pseudorapidity all one has to do is to define how the
initial density and shape of the system changes along the beam
direction.  This, however, would lead to a very counterintuitive
initial shape of the system. The observed multiplicity stays
approximatively constant within three units of
pseudorapidity~\cite{Phobos} which -- assuming that the longitudinal
flow is approximatively boost invariant -- would mean that the initial
entropy per unit of flow rapidity is approximatively constant within
three units of rapidity. On the other hand the elliptic flow data
peaks at midrapidity~\cite{phobos}.  Combined with the requirement of
constant entropy, this would mean that the initial system shape is
most asymmetric at midrapidity and becomes more cylindrical when
fragmentation regions are approached. It is difficult to imagine a
physical process leading to this kind of distribution. So far
hydrodynamical calculations~\cite{Hirano:2001,Hirano:2002} have not
tried to find the initial shape to fit the data by trial and error,
but have instead tried to formulate an intuitive parametrization for
the initial shape and calculated the elliptic flow based on
that. These parametrizations either assume the shape not to change or
to become more asymmetric towards the fragmentation regions. As a
result the calculated elliptic flow fits the data only around
midrapidity where $|\eta| < 1$ and stays almost constant in a large
region of pseudorapidity.

There is considerable freedom in choosing the initial state in
hydrodynamics and therefore it is premature to draw final conclusions
as to whether the observed pseudorapidity dependence of elliptic flow
really means a deviation from hydrodynamical behaviour. There may be a
reasonable way to tune the initial conditions to reproduce the data
which we have not thought about yet.

\section{Summary and outlook}
   \label{summary}

Hydrodynamical models have been surprisingly successful in explaining
the flow data obtained at RHIC. The differential elliptic flow of
charged particles is reproduced up to $p_T=1.5$ GeV, its centrality
dependence works below impact parameter $b\approx 6$ fm and the
differential elliptic flow of identified particles shows the mass
ordering predicted by hydrodynamical calculations. That a model which
is based on thermal averages is able to reproduce details at the
5-10\% level is remarkable. This success, however, leaves us two
complementary puzzles to solve: If the system does not thermalize, why
is the observed elliptic flow so close to hydrodynamical elliptic
flow? Or, if the system thermalizes, what is the mechanism responsible
for it and how does one describe the thermalization process? At the
time of this writing, there are no answers to either of these
questions.

The usefulness of flow studies and hydrodynamical models is not
limited to deducing whether the collision system thermalized or
not. As described in the previous chapter, flow is affected by the
freeze-out temperature, the equation of state and the initial
state. Thus it is possible to obtain information on all of them by
studying flow, but, on the other hand, the complicated dependencies
makes it challenging to constrain them individually. This requires
careful simulations where one tries to reproduce as large an amount of
the data as is possible and where the effects of all variables are
taken into account. So far it is possible to say that the flow data
favours an equation of state with a phase transition and initial state
with densities well above the phase transition temperature. However,
there are also open questions like the pseudorapidity dependence of
flow and the HBT radii which must be solved before conclusions can be
drawn.

\section*{Acknowledgements}
I am indebted to many colleagues for fruitful discussions and debates.
Especially I want to thank Ulrich Heinz, Peter Kolb and Denes Molnar
for comments and help. This work was supported by the US
Department of Energy grant DE-FG02-87ER40328.

\end{document}